\documentclass[%
 preprint,
 amsmath,amssymb,
 aps,
floatfix,
]{revtex4-2}

\usepackage{graphicx}
\usepackage{dcolumn}
\usepackage{bm}
\usepackage{physics}
\usepackage{color}
\usepackage{subfigure}

\begin{document}

\preprint{APS/123-QED}

\title{The chain length of anisotropic paramagnetic particles in a rotating field}

\author{Jānis Užulis}
\author{Jānis Cīmurs}%
 \email{janis.cimurs@lu.lv}
 \homepage{\href{https://mmml.lu.lv/}{mmml.lu.lv}}
\affiliation{%
 Laboratory of Magnetic Soft Materials\\
 Faculty of Physics, Mathematics and Optometry\\
 University of Latvia\\
 Jelgavas iela 3-530, Riga, Latvia, LV-1004
}%

\date{\today}

\begin{abstract}
In this article the maximal length of a chain of paramagnetic particles with magnetic anisotropy in a rotating magnetic field is studied.
The theory of paramagnetic particle chains usually assumes that the particles are magnetically isotropic and do not rotate in a rotating field. In experiments it is seen that spherical paramagnetic particles rotate, which can be explained by small magnetic anisotropy. In this article, the maximal chain length is calculated for paramagnetic particles with magnetic anisotropy in a rotating magnetic field. Results show that the maximal chain length as a function of field frequency has the same trend for isotropic magnetic particles and particles with magnetic anisotropy if the field frequency is much higher or much lower than the critical frequency of an individual particle.
blue Initially randomly distributed particles will form chains that will collide and exchange with particles till they obtain a typical chain length.The typical chain length of a small cluster is shorter than the maximal chain length of an isolated chain for the same field frequency. The distribution of chain lengths in a small cluster of chains is narrower for particles with higher magnetic anisotropy. Due to the narrower distribution of chain lengths, particles with magnetic anisotropy can suit better for mass-production.
This article will show how magnetic anisotropy parameters of paramagnetic particles influence chain length of chains which form in a rotating magnetic field.

\end{abstract}

\maketitle


\section{Introduction}

It is shown that paramagnetic particles in a rotating magnetic field become attractive \cite{yigit2020,Massana-Cid2019}. If the field frequency is small, particles form chains \cite{yigit2020, Han2020}, where the chain length depends on the field frequency \cite{melle2003,gao2012,vazquez1017,devi2015} and fluid and particle parameters.

Magnetic particles chains can be used for cargo transport \cite{Massana-Cid2019, Lo2019}. The advantage of the magnetic particles in cargo transport is that the motion can be fully controlled from the outside using a pattern of an external magnetic field.
Functionalized magnetic particles can be linked together creating magnetic filament \cite{Cebers2004,biswal2004} which can be used as magnetic swimmer \cite{tierno2014}. For these applications, the possibility of creating a large amount of chains with well defined length is advantageous.

Magnetic fluid, which consists of superparamagnetic particles suspended in a liquid, also exhibits a similar formation of chains in a rotating magnetic field \cite{stikuts2020,Chen2015}. Despite the fact that in a phase separated magnetic fluid surface tension and thermal motion of the particles play an important role, the behaviour of magnetic fluid in a rotating field is similar to the behaviour of a small cluster of magnetic particles. In slow rotating magnetic fields, magnetic fluid forms elongated drops with narrow distribution of lengths \cite{stikuts2020}.

The theory of spherical paramagnetic particles in a rotating magnetic field usually does not take into account magnetic anisotropy of particles, therefore particles are assumed non-rotating. In reality it can be seen that paramagnetic particles rotate in a rotating field \cite{Massana-Cid2019,martinez2015}. In \cite{Massana-Cid2019,martinez2015}, rotation of spherical paramagnetic particles is explained by finite relaxation time of magnetic moment. The rotation of the paramagnetic particles can be explained also by magnetic anisotropy of the susceptibility of the particles.

In this article, chain formation of paramagnetic particles with anisotropy in a rotating magnetic field is investigated. The critical chain length is compared to theory for chains of isotropic paramagnetic particles to see if measurable difference can be observed. Also, the hypothesis that particles with magnetic anisotropy have more narrow chain length distribution in initially randomly distributed cluster of particles is tested. The results and conclusions show advantages for using paramagnetic particles with anisotropy to form paramagnetic chains.

\section{Model}
\label{sec:model}
\begin{figure}[htbp]
\includegraphics[
 width=240pt,
  keepaspectratio]{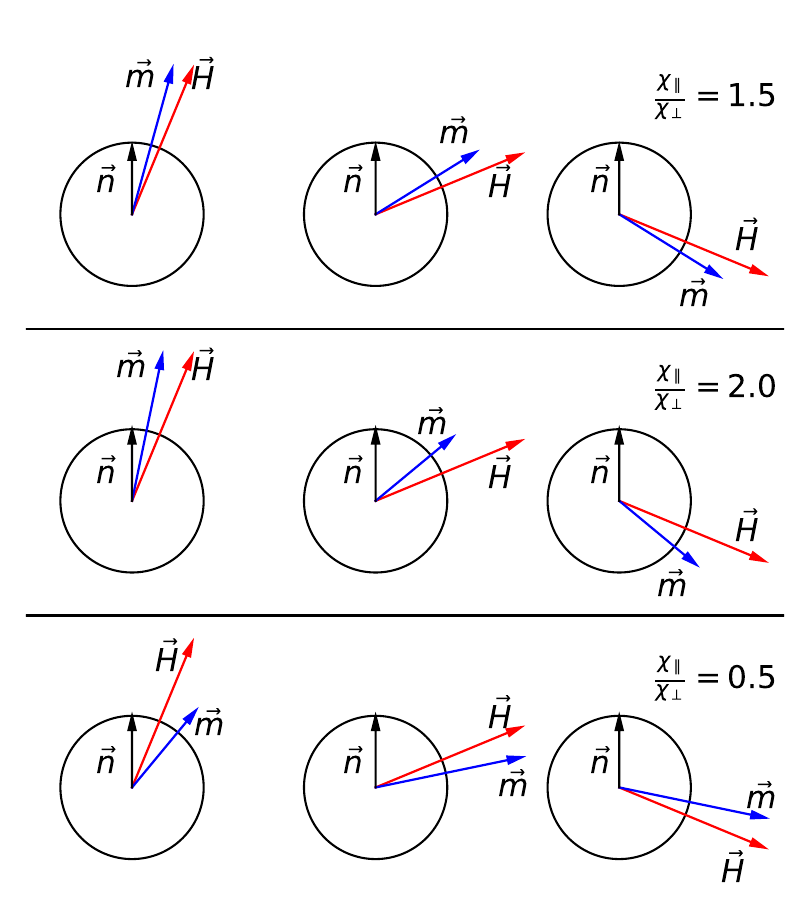}
  
\caption{ \label{fig:anisotr_ilustr} Magnetic moment $\vec m$ direction depending on the magnetic field $\vec H$ direction. The magnetization $|\vec m|$ is normalized to the field strength $|\vec H|$ for each anisotropy case. In upper two examples $\vec n$ is an easy magnetization axis ($\frac{\chi_\|}{\chi_\perp}=2$, $\frac{\chi_\|}{\chi_\perp}=1.5$) and in the lower one example $\vec n$ is a hard magnetization axis ($\frac{\chi_\|}{\chi_\perp}=0.5$). The direction of the magnetization $\vec m$ and the external field $\vec H$ direction coincide when the direction of $\vec H$ is along the anisotropy axis $\vec n$ or perpendicular to it.
}
\end{figure}

In this article, we will describe  paramagnetic particles with spherical form and with uniaxial magnetic anisotropy. The magnetic anisotropy could be caused by the crystallographic structure \cite{aharoni2000} or by the fact that particles are not ideally spherical or a particle could be a magnetic rod in a spherical shell \cite{Mukhtar2020} or due to magnetic Janus particles, where some regions have different magnetic susceptibility  \cite{yammine2020}. When the particles are subjected to a weak magnetic field $\vec H$, the magnetic dipole moment $\vec m_i$ is induced in each particle. The dipole moment of $i$-th particle can be calculated using a linear particle magnetization and field relation:
\begin{equation}\label{eq:m}
  \vec m_i=V_m\left(\chi_\perp \vec H_i +\Delta\chi(\vec n_i\cdot\vec H_i)\vec n_i\right)\text{ ,}
\end{equation}
where $V_m$ is the magnetic volume of the particle, $\vec H_i$ is the magnetic field at the point of the particle, $\vec n_i$ is the unit vector in the direction of the anisotropy axis of the particle,  $\Delta\chi=\chi_\|-\chi_\perp$ is difference of the magnetic susceptibilities, where $\chi_\|$ and $\chi_\perp$ are magnetic susceptibility of the particle in the direction of anisotropy axis and perpendicular to it respectively. For ellipsoidal paramagnetic particles, susceptibilities $\chi_\perp$ and $\chi_\|$ are related to material susceptibility $\chi$, as shown by Stoner and Osborn \cite{Stoner1945,Osborn1945}.
$\chi_{\|,\perp}=\chi_i/(1-N_{\|,\perp} \chi_i)$, where $\chi_i$ is material susceptibility in the given direction and $N_{\|,\perp}$ is a demagnetization factor which depends on the shape of the particle.
For example, an isotropic spherical particle with large magnetic susceptibility has $N_{\|}=N_\perp=1/3$ and $\chi_\|=\chi_\perp=3$. Parameter $\frac{\chi_\|}{\chi_\perp}$ is used to characterize the anisotropy of the particle in this article.
It is assumed that magnetic relaxation time is much shorter than the rotation period of the external magnetic field. As a result, the magnetic moment calculated by equation \eqref{eq:m} does not depend on the rotation frequency. The direction of magnetization $\vec m$ is illustrated in Figure \ref{fig:anisotr_ilustr} for different external fields $\vec H$ and anisotropy axis $\vec n$ configurations. 

Figure \ref{fig:anisotr_ilustr} shows that the magnetization $\vec m$ is closer to the anisotropy axis if $\chi_\|>\chi_\perp$ and the anisotropy axis is an easy axis. If $\chi_\perp>\chi_\|$ the anisotropy axis is a hard axis and magnetization of the particle is closer to the equator of the particle.

In the model these particles are dispersed in liquid with viscosity $\eta$. 
Each particle rotates due to torque $\vec \tau_i=\mu_0 \vec m_i\times\vec H_i$, which is compensated by viscous drag. The rotation of the particle can be calculated as \cite{Cimurs2019}:
\begin{equation}
\label{eq:dn/dt}
  \dfrac{\dd \vec n}{\dd t}=\dfrac{\mu_0 V_m\Delta\chi}{\xi_r}\left(\vec n\cdot\vec H_i\right)\left[\vec H_i-\left(\vec n\cdot\vec H_i\right)\vec n\right]\text{ ,}
\end{equation}
where $\vec n$ is the magnetic anisotropy axis, which is fixed to the particle, $\mu_0$ is the magnetic constant (vacuum permeability), $\xi_r=8\pi\eta R^3$ is the rotational drag coefficient, where $R$ is the hydrodynamic radius of the particle and $\eta$ is viscosity of the liquid. It is assumed that viscosity dominates over inertia and Reynolds number is below unity and inertial terms can be neglected.

The magnetic field at the point of the particle $\vec H_i$ is a sum of the external field $\vec H$ and the field produced by all other particles. $j$-th particle at the point of $i$-th particle produces field:
\begin{equation}\label{eq:H}
\displaystyle \vec {H}_{ij} (\vec {r}_{ij} )=\frac {1}{4\pi}\left({\frac {3\vec {r}_{ij} (\vec {m_j} \cdot \vec {r}_{ij} )}{r_{ij}^{5}}}-{\frac {\vec {m_j} }{r_{ij}^{3}}}\right)\text{ ,}
\end{equation}
where $\vec r_{ij}=\vec r_i-\vec r_j$ is distance vector from the point of $j$-th particle to the point of $i$-th particle. The magnetic field at the point of the $i$-th particle is $\vec H_i=\vec H+\sum\limits_j \vec H_{ij}$

Since the field produced by particles is not homogeneous, particles also move. The force exerted by $j$-th particle on the $i$-th particle can be calculated as
\begin{equation}\label{eq:F}
\begin{split}
\displaystyle \vec {F}_{ij,M}&={\dfrac {3\mu _{0}}{4\pi r_{ij}^{5}}}\bigg[(\vec {m} _{i}\cdot \vec {r}_{ij} )\vec {m} _{j}+(\vec {m} _{j}\cdot \vec {r}_{ij} )\vec {m} _{i}\\&+(\vec {m} _{i}\cdot \vec {m} _{j})\vec {r}_{ij} -{\dfrac {5(\vec {m} _{i}\cdot \vec {r}_{ij} )(\vec {m} _{j}\cdot \vec {r}_{ij} )}{r^{2}_{ij}}}\vec {r}_{ij} \bigg]\text{ .}
\end{split}
\end{equation}

The non-overlap of the particles is fulfilled using the repulsive part of Lenard-Jones potential. The repulsive force is:
\begin{equation}
\label{eq:F_rep}
\vec F_{ij,R}=\begin{cases}
-G\left(
\frac{1}{r_{ij}^{13}}-\frac{r_{ij}-2R}{r_{max}-2R}\frac{1}{ r_{max}^{13}}
\right ) \frac{\vec r_{ij}}{r_{ij}} \text{ if } r_{ij}\leq r_{max} \\
0 \text{ if }  r_{ij}>r_{max} \\
\end{cases}\text{.}
\end{equation}
The repulsive force is proportional to the particle distance in power of $-13$ as obtained from the Lenard-Jones potential repulsive part.
Constant $G$ is chosen so that the maximal attractive magnetic force $\vec F_{ij,M}$ between two particles at the distance of $|r_{ij}|=2R$ between particle centers is compensated by repulsive force $\vec F_{ij,R}$ and $\vec F_{ij,M,max}+\vec F_{ij,R}=0$.
The maximal attractive magnetic force is obtained in a situation where $\vec n_i\|\vec H_i$ and $\chi_\|>\chi_\perp$ and $\vec r_{ij}\|\vec m_i\|\vec m_j$. The maximal magnetic force is  $\vec F_{ij,M,max}=\frac{3V_m^2\chi_\|^2H_i^2\mu_0}{32\pi R^4}$ and consequently $G=F_{ij,M,max}(2R)^{13}$.
The parameter $r_{max}$ is a bit larger than $2R$ and determines the softness of the particle. The equation part $\frac{r_{ij}-2R}{r_{max}-2R} \frac{1}{ r_{max}^{13}}$ makes the repulsive force 0 at the particle distance $r_{ij}=r_{max}$ and makes the repulsion function without steps, to avoid numerical instabilities.

The sum of all forces, the magnetic force \eqref{eq:F} and the repulsive force \eqref{eq:F_rep}, will cause the particle to move.
In the non-inertial limit the driving force is compensated by drag force for spherical particle $\vec F_D=-6\pi\eta R \vec v$, where $\vec v$ is velocity of the particle. Velocity of the $i$-th particle is calculated as:
\begin{equation}
\label{eq:v_i}
\vec v_i=\frac{1}{6\pi \eta R}\sum\limits_j \vec F_{ij},
\end{equation}
where $\vec F_{ij}=\vec {F}_{ij,M}+\vec {F}_{ij,R}$ is a sum of all the forces.

Magnetic particles are subjected by an external rotating magnetic field: $\vec H=H (\cos(\omega_H t), \sin(\omega_H t),0)$, where $\omega_H$ is a field rotation frequency. 

In this article thermal fluctuations are omitted. Thermal fluctuations are irrelevant if the displacement due to thermal fluctuations is much smaller than the displacement due to magnetic forces and torques. We can rewrite this statement as $\langle \vec r^2\rangle\gg D_T t$ and $\langle \vec \theta^2\rangle\gg D_R t$, where $\langle \vec r^2\rangle$ is displacement and $\langle \vec \theta^2\rangle$ is rotation of the particle in characteristic time $t$, $D_T$ and $D_R$ are translation and rotation diffusion constants respectively. If we take $1/\omega_H$ as the characteristic time and $R$ as the characteristic distance. we can obtain that the radius of the particle should fulfill $R^3\gg (k_B T)/(4\pi\eta\omega_H)$. The typical values of $\sqrt[3]{(k_B T)/(4\pi\eta\omega_H)}$ are in the range $[100 nm, 1\mu m]$. So the particle size should be at least 10 times larger than the value of $\sqrt[3]{(k_B T)/(4\pi\eta\omega_H)}$ which would inherit that $R^3$ is 1000 times larger than $(k_B T)/(4\pi\eta\omega_H)$.

\section{Theory}


The theory by Melle et al. \cite{melle2003} shows that isotropic particles in a slow rotating magnetic field form chains with length proportional to $\frac{1}{\sqrt{\omega_H}}$. In this article we will show that if a particle has an anisotropy of magnetic susceptibility, then the chain length in the rotating magnetic field deviates from the trend $\frac{1}{\sqrt{\omega_H}}$ if the frequency $\omega_H$ is close to the critical frequency of a single particle $\omega_{C,1}$, where
\[\omega_{C,1}=\frac{\mu_0V_m\Delta\chi H^2}{2\xi_r}=\frac{\mu_0V_m\Delta\chi H^2}{16\pi\eta R^3}\] and $H$ is the field strength of the external magnetic field. 

We call $\omega_{C,1}$ the critical frequency of a single particle because an isolated paramagnetic particle rotates synchronously with the field if $\omega_H<\omega_{C,1}$ and asynchronously if $\omega_H>\omega_{C,1}$.\cite{Cebers2003}
If the particles magnetic anisotropy axis is a hard axis ($\chi_\|<\chi_\perp$), then the anisotropy axis lies perpendicularly to the plane of the rotation of the magnetic field and the directions of $\vec m$ and $\vec H$ coincide. The rotation of the particle is no longer happening and the chain formation is the same as for isotropic particles where chain length, as shown by Melle et al \cite{melle2003}, is proportional to $\frac{1}{\sqrt{\omega_H}}$.

For example, a spheroidal particle whose symmetry axis is 1\% longer than the other two axis and which is made of material with magnetic susceptibility $\xi=7$ (in SI units), in water, in magnetic field with strength $H=1.6 kA/m$, will have the critical frequency $\omega_{C,1}=11 rad/s$. The critical frequency does not depend on the size of the particle but on the shape of the magnetic core. The critical frequency is highly effected by the field strength and can also be manipulated by viscosity of the fluid

Further in this article we will use the term critical frequency $\omega_{C,N}$ to emphasize difference between the current rotation frequency $\omega_H$ and the maximal possible field frequency for the given chain $\omega_{C,N}$. If the external rotation frequency $\omega_H<\omega_{C,N}$, isolated chain of length $N$ rotates with the field, but if $\omega_H>\omega_{C,N}$, the isolated chain breaks. As a result, the theory of Melle \cite{melle2003} becomes $N\propto\frac{1}{\sqrt{\omega_{C,N}}}$

\subsection{Slow field}
For particles with a magnetic anisotropy easy axis ($\chi_\|>\chi_\perp$), the anisotropy axis is in the plane of the rotating field and the magnetic moment of the particle depends on the angle between the external field $\vec H$ and the anisotropy axis of the particle $\vec n$.
In a slow rotating magnetic field $\omega_H<\omega_{C,1}$ the particle follows the field and the magnetic moment of the particle is constant: 
\[\abs{\vec m}=V_m \sqrt{\chi_\perp^2\vec H^2+2\chi_\perp\Delta\chi \left(\vec n\cdot\vec H\right)^2+\Delta\chi^2\left(\vec n\cdot\vec H\right)^2}\text{ ,}\]
where \cite{Cimurs2019}
\[\left(\vec n\cdot\vec H\right)^2=\frac{H^2}{2}\left(1+\sqrt{1-\left(\frac{\omega_H}{\omega_{C,1}}\right)^2}\right)\text{ .}\]
From the theory of Melle et al. \cite{melle2003} chain will break if the angle between the chain and the external magnetic field exceeds the critical value $\frac{\pi}{4}$. In other words, the chain will break if hydrodynamic drag $6\pi \eta R^2\omega_{C,N} N^2$  overcomes magnetic force $\frac{3\mu_0}{4\pi (2 R)^4}|\vec m|^2$.
Putting all together, it gives a formula connecting the chain length $N$ with the critical frequency $\omega_{C,N}$ in a slow rotating field:
\begin{equation}\label{eq:N_theor}
  N=\sqrt{\frac{V_m \omega_{C,1}}{\Delta\chi 16 \pi R^3 \omega_{C,N}}}\sqrt{\chi_\perp^2+\chi_\|^2+(\chi_\|^2-\chi_\perp^2)\sqrt{1-\frac{\omega_{C,N}^2}{\omega_{C,1}^2}}}
\end{equation}

It can be shown that the chain length $N$ deviates from the trend $N\propto\frac{1}{\sqrt{\omega_{C,N}}}$ only for values of $\frac{\omega_{C,N}}{\omega_{C,1}}$ close to unity. If the rotation frequency is small ($\omega_{C,N}\ll\omega_{C,1}$), the equation \eqref{eq:N_theor} becomes
\begin{equation}\label{eq:N_slow}
  N=\sqrt{\frac{V_m\chi_\|^2}{8\pi\Delta\chi R^3}}\sqrt{\frac{\omega_{C,1}}{\omega_{C,N}}}=\sqrt{\frac{\mu_0 V_m^2 H^2\chi_\|^2}{16\pi\xi_r R^3}}\sqrt{\frac{1}{\omega_{C,N}}}\text{ ,}
\end{equation}
which is proportionality $N\propto\frac{1}{\sqrt{\omega_{C,N}}}$.
For particles which are almost spherical, the critical rotation frequency $\omega_{C,1}$ can become extremely small and frequencies $\omega_{C,N}$ corresponding to $N>1$ are much higher than the frequency $\omega_{C,1}$ and the formula \eqref{eq:N_theor} is not suitable. 

\subsection{Fast field}

Particles, which are almost isotropic, have low critical field frequency $\omega_{C,1}$ and do not follow the field. This regime is called asynchronous rotation because the angle between the external field $\vec H$ and the particle direction $\vec n$ changes. Numerical simulations show that, if the field frequency is large enough, particles synchronise with each other.

The force between two neighbouring synchronised particles would be:
\begin{equation}\label{eq:F_ij_eq}
    \vec F_{ij}= \dfrac {3\mu _{0}}{4\pi (2R)^4}\left[2(\vec m\cdot \hat r )\vec m+\vec m^2\hat r -5(\vec m \cdot \hat r)^2\hat r \right]\text{ ,}
\end{equation}
where attractive radial force is
\[
  F_r=\dfrac {3\mu _{0}}{4\pi (2R)^4}\left[\vec m^2 -3(\vec m \cdot \hat r)^2 \right]
\]
and the force which makes particles to rotate along common center is
\[
  \vec F_\theta= \dfrac {3\mu _{0}}{2\pi (2R)^4}(\vec m\cdot \hat r )(\vec m\cdot\hat\theta)\text{ ,}
\]
where $\hat r$ is a unit vector in the direction to the other particle and $\hat \theta$ is a unit vector perpendicular to $\hat r$ and in the direction of the rotation. Using Eq. \eqref{eq:m} and introducing angle $\theta$ between $\vec H$ and $\vec r$ and angle $\varphi$ between $\vec n$ and $\hat r$ gives relations for magnetic moment $\vec m$:
\[
  \hat r\cdot \vec m
    =V_m H\Big(\chi_\perp \cos\theta+\Delta\chi \cos\varphi\cos(\theta-\varphi)\Big)
\]
\[
  \vec m^2
    =V_m^2 H^2\Big(\chi_\perp^2 +(\chi_\|^2-\chi_\perp^2)\cos^2(\theta-\varphi)\Big)
\]
\[
  \vec m\cdot\hat\theta
    =V_m H\Big(-\chi_\perp \sin\theta+\Delta\chi \sin\varphi \cos(\theta-\varphi)\Big)
\]

It will be assumed that in high frequency $\theta$ changes slowly and $\varphi$ increases linearly with time. This provides a possibility to average trigonometric functions over linearly increasing $\varphi$. The average inter-particle force will read
\[
  \langle F_r\rangle=-\dfrac{3\mu _{0}V_m^2 H^2}{32\pi (2R)^4}\left(2(\chi_\perp^2+\chi_\|^2)+3(\chi_\perp+\chi_\|)^2\cos(2\vartheta)\right)
\]
\[
  \langle F_\theta\rangle=-\dfrac{3\mu _{0}V_m^2 H^2}{16\pi (2R)^4}(\chi_\perp+\chi_\|)^2\sin(2\vartheta)
\]
For reasonable $\frac{\chi_\|}{\chi_\perp}$, the average force is close to an isotropic situation, where $\chi_\|=\chi_\perp$.

The conclusion in high frequency range is similar to the conclusion for isotropic particles \cite{melle2003}.
In an anisotropic situation the angle $\theta$, at which $\langle F_r\rangle$ becomes repulsive, is larger than the magic angle, for any $\frac{\chi_\|}{\chi_\perp}$ value. 
Therefore, breakage of the chain will happen when friction force will overcome maximal $\langle F_\theta\rangle$. As a result, the maximal chain length in a high frequency range is:
\begin{equation}\label{eq:N_asinh}
  N=\sqrt{\frac{\mu_0 V_m^2H^2 (\chi_\|+\chi_\perp)^2}{64\pi R^3 \xi_r}}\sqrt{\frac{1}{\omega_{C,N}}}
\end{equation}
As can be seen, in high frequencies, the chain length is also proportional to  $\frac{1}{\sqrt{\omega_C,N}}$. The proportionality constant in the fast field is different from the proportionality constant found in the slow field. The proportionality constant is higher for the slow field than for the fast field if $\chi_\|>\chi_\perp$.

When the field frequency gets close to the critical frequency, the given equation of chain length Eq. \ref{eq:N_asinh} will not work. To calculate a more precise relation, the back and forth rotation of each particle should be considered in the derivation, which gives field frequency dependent mean values for trigonometric functions of $\varphi$. For example, mean value $\langle\sin(2\varphi)\rangle=\frac{\omega_H-\sqrt{\omega_H^2-\omega_{C,1}^2}}{\omega_{C,1}}$ \cite{Cebers2003}. In addition, in asynchronous rotation of the particles, the assumption that the angle $\theta$ between the chain and the field does not change does not hold. The fact is that the angle $\theta$ changes do not allow to obtain maximal chain length in the closed form in medium frequencies in the asynchronous regime.

\subsection{Long chain correction}
The results obtained so far were based on the assumption that farther particles do not influence the magnetic field of middle particles. This is not true even for two particles. For longer chains, the influence of all particles should be considered. Usually in theory an infinitely long chain is considered. In an infinitely long  straight chain the force and the torque on each particle can be calculated, and they involve two angles ($\varphi$ and $\theta$) in different trigonometric combinations. This does not allow to obtain a solution in the closed form even for low field frequencies. In simulations, it can be seen that the chain is not straight but S-shape. As a result, further analysis of chain lengths is based on simulations results.

\section{Results}
The model described in section \ref{sec:model} was implemented in a C++ computer program, to do numerical simulations. Translation motion Eq.(\ref{eq:v_i}) and rotation motion Eq.(\ref{eq:dn/dt}) for each particle is being calculated by solving first order ordinary differential equation system of $6N$ equations, where $N$ is the number of particles. GNU Scientific Library \cite{GNU_ode}, Runge-Kutta 4th order method with adaptive time step were used to solve the differential equation system.

\subsection{Soft VS hard particles}

\begin{figure}[htbp]
\centering
\subfigure{%
\includegraphics[
  width=240pt]{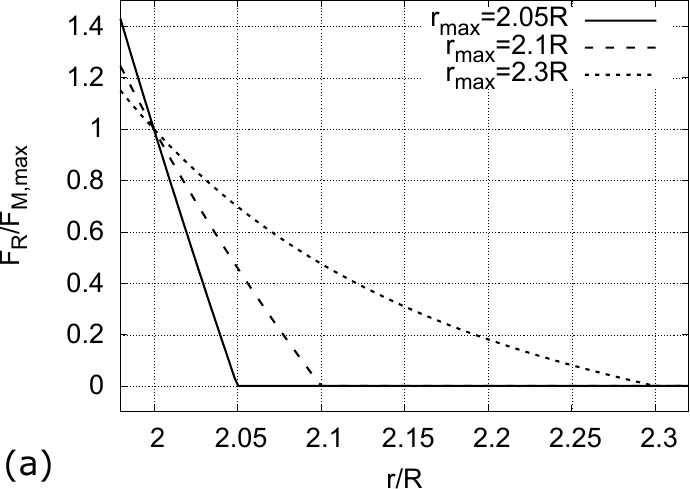}
\label{fig:F_R}}

\subfigure{%
\includegraphics
 [ width=240pt]
  {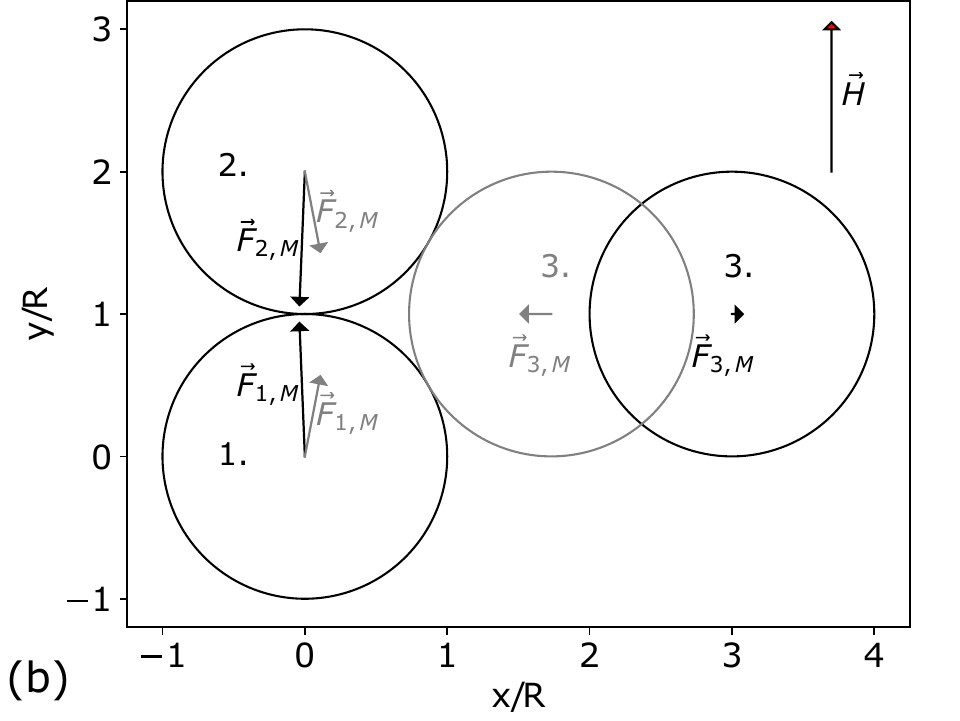}
\label{fig:3lodes}}
\caption{(a) Repulsive force Eq. \eqref{eq:F_rep} $F_R$ depending on particle distance $r$, with different parameter $r_{max}$, normalized to maximal magnetic force of two particles at distance $2R$. Steepness of the repulsive force function $F_R$ (the softness of the particle) can be regulated by choosing parameter $r_{max}$. (b) A two particle chain and a situation where the third particle is at a distance where the force is repulsive (particle and forces in black) and a situation where the third particle is at a distance where the force is attractive to the chain (particle and forces in gray). Sizes of the force vectors are drawn in a scale where $F_{i,M,max}=1$.}
\end{figure}

By choosing parameter $r_{max}$ value in Eq.(\ref{eq:F_rep}), steepness of the repulsion force function can be regulated. The smaller $r_{max}$ value we choose, the steeper the growth of the repulsive force, as can be seen in Fig. \ref{fig:F_R}.

If there is a chain of two particles (a dimer) and the third particle comes closer to the dimer, like illustrated in Fig.\ref{fig:3lodes}, the magnetic force acting on the third particle is repulsive if its position $x>2R$.
If the third particle position $x<2R$, then the magnetic force on the third particle is attractive to the dimer and they form a trimer. The distance between the center of the third particle and the line of centers of dimer particles, where the force changes direction, is $\sqrt{5}R$.
If the value of $r_{max}$ is larger than $\sqrt{5}R$, then in a cluster of particles the probability for particles to form a trimer is small and these three particles will form a chain structure.
The increase of $r_{max}$ not only increases the distance of the repulsive force but also makes particles softer.
The particles close to each other amplify the field and the repulsive force at the distance 2R does not compensate the attractive magnetic force completely, and the force equilibrium distance is at a distance a bit less than 2R, which makes the force acting on the third particle flip direction at a distance a bit less than $\sqrt{5}R$.
If larger $r_{max}$ value is used, the particles are softer than with small $r_{max}$ value.


In the article \cite{2003.03008_SPION} the order of the chain in the external field is being regulated by salt concentration in the liquid. The higher the salt concentration, the less disordered the chain gets. In our model similar structural effects are observed by changing $r_{max}$ value. Higher $r_{max}$ value corresponds to higher salt concentration. Despite the fact that changing salt concentration gives qualitatively similar results as changing $r_{max}$ value, the dynamics of such charged colloids could differ from the results obtained in this article. Movement of the charged particle would lead to a shifted ionic layer around the particle that creates electrostatic interactions not included in the model discussed in this article.

\subsection{The critical length of a single chain}
At the beginning, $N$ particles are placed in a straight chain along the direction of a stationary field. At this phase particles obtain equilibrium distance between each other. Then the field starts to rotate with a certain frequency $\omega_H$. If the frequency is higher than the critical frequency $\omega_{C,N}$, the chain breaks after some periods, but, if the frequency is below the critical - the chain does not break. There is a very well observable boundary between the frequency where the chain breaks in a few periods and the frequency where the chain does not break in even a couple of hundred periods.

\begin{figure}[htbp]
\includegraphics
 [width=240pt]
  {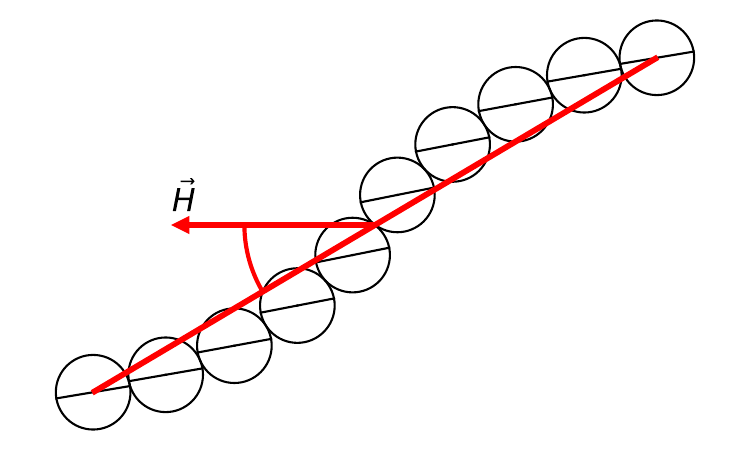}
\caption{ \label{fig:kede_lenkis} S-shaped chain at the critical field rotation frequency $\omega_{C,10}$. The angle between the line of the end particles of the chain and the external field $\vec H$ are being monitored to detect breaking of the chain. The field rotates clockwise. The easy magnetization axis is shown as a black line in a particle.
}
\end{figure}

The angle between the line of the end particles of the chain and the external field  (Fig.\ref{fig:kede_lenkis}) is being monitored to detect when the chain breaks. If the angle gets larger than $\frac{\pi}{2}$, then the chain is definitely broken. By monitoring this angle even breaking of a two particle chain can be detected.

\begin{figure}[htbp]
\includegraphics
 [width=240pt]
  {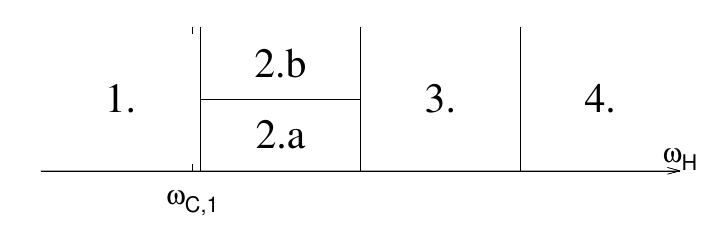}
\caption{ \label{fig:modes} Chain rotation mode classification depending on the field rotation frequency $\omega_H$ if the field rotation frequency is close to the critical frequency $\omega_{C,N}$.
\textbf{1.} Synchronous mode (each particle rotates synchronously with the field).
\textbf{2.a} Both end particles rotate asynchronously with the field.
\textbf{2.b} Middle particle of the chain rotates asynchronously with the field.
\textbf{3.} More than two particles rotate asynchronously with the field and their rotation frequencies do not match.
\textbf{4.} All particles rotate slower than the field and they rotate synchronously with each other.
Examples of all modes in video format can be found in supplementary materials. The positions of the boundaries between the modes depend on the particle and field properties.
}

\end{figure}

If the field rotation frequency $\omega_H$ is slightly smaller than the critical rotation frequency $\omega_{C,N}$, for a given chain, 4 structurally different rotation regimes can be observed.
If $\omega_H<\omega_{C,1}$, then each particle rotates synchronously with the field (mode 1. Fig.\ref{fig:modes}).
If $\omega_H$ is above $\omega_{C,1}$, then two scenarios can be observed.
In the first scenario, the end particles of the chain start to lag behind the field and their rotation becomes asynchronous with the field, but the middle particles rotate synchronously with the field (mode 2.a Fig. \ref{fig:modes}).
In the second scenario, the middle particle of the chain starts to lag the field (mode 2.b Fig. \ref{fig:modes}).
If the frequency is increased further, then more than two particles rotate asynchronously with the field, and the rotation of the middle particles is faster than the rotation of the end particles (mode 3. Fig. \ref{fig:modes}).
If the field frequency is increased even more, then the rotation frequency for all particles becomes so small that it can be assumed that the particles synchronize  (mode 4. Fig. \ref{fig:modes}). Visual representation of all modes can be found in the supplementary video.

\begin{figure}[htbp]
\includegraphics
 [width=240pt]
  {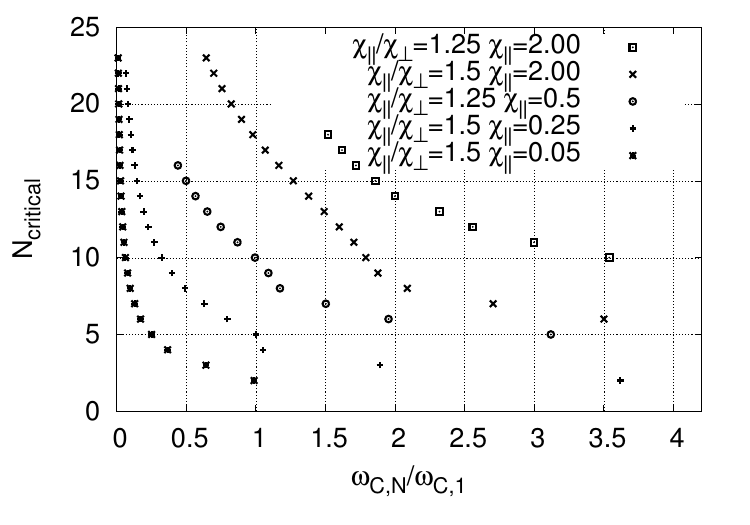}
\caption{ \label{fig:graph} Chain length $N$ and its critical field rotation frequency $\frac{\omega_{C,N}}{\omega_{C,1}}$ for chains with different magnetic anisotropy properties. Transitions between chain rotation modes are visible like step or change of steepness of the graph. 
}
\end{figure}

Fig.\ref{fig:graph} shows the critical chain length $N$ dependence on the field frequency for particles with different magnetic anisotropy properties.
For particles with $\frac{\chi_\|}{\chi_\perp}=1.5$ and $\chi_\|=2$ there is a transition between mode 1. and mode 2.a for a chain length of 13 and 12 particles (a 13 particles chain at its critical frequency $\omega_{C,13}=1.49$ rotates in mode 1 and a 12 particle chain at its critical frequency $\omega_{C,12}=1.6$ rotates in mode 2.a).
This mode change is not particularly pronounced in Fig. \ref{fig:graph} if the chain is long, but, for shorter chains, like in example $\frac{\chi_\|}{\chi_\perp}=1.5$ and $\chi_\|=0.25$, the mode transition is visible as a step in a plot between chain length of $5$ and $4$ particles.

For particles $\frac{\chi_\|}{\chi_\perp}=1.25$ and $\chi_\|=0.5$ there is a transition from mode 1. to mode 2.b at chain length of 10 to 9, and an 8 particle chain rotates in mode 3. 
The transition between mode 2.a and mode 3. is visible in example $\frac{\chi_\|}{\chi_\perp}=1.5$ and $\chi_\|=2$ for a chain length of 10 and 9 particles. Steepness of the plot changes as the mode changes.

\begin{figure}[htbp]
\subfigure{\includegraphics
 [width=240pt]
  {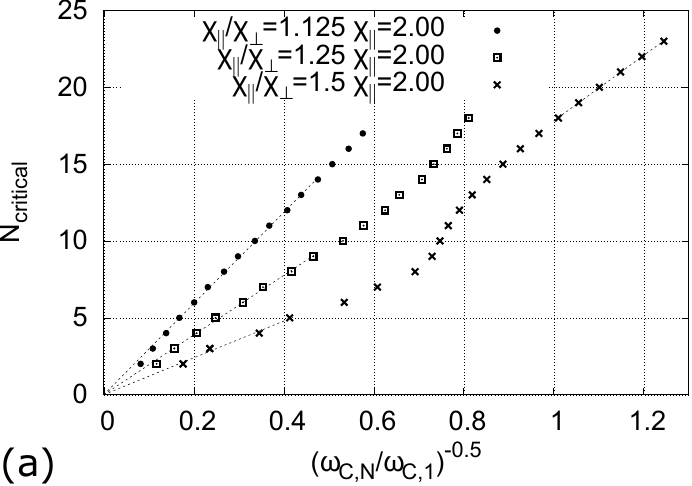} \label{fig:mode4fit}}
\subfigure{\includegraphics
 [width=240pt]
  {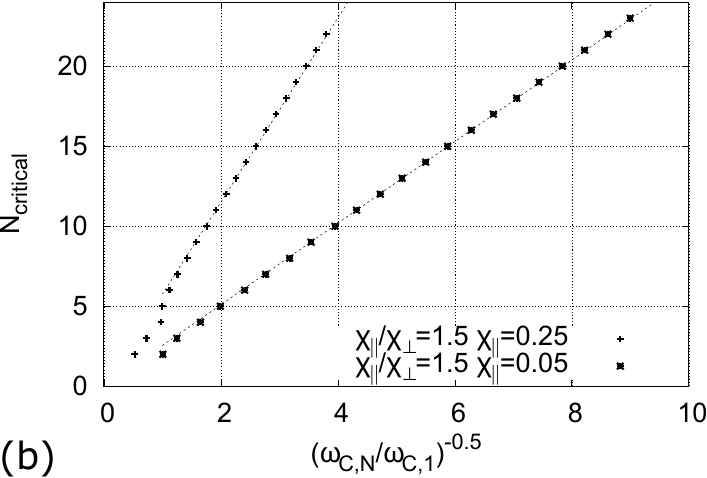} \label{fig:mode1fit}}
\caption{The critical chain length $N_{critical}$ as a function of $\sqrt{\frac{\omega_{C,1}}{\omega_{C,N}}}$ for chains with different magnetic anisotropy properties. Linear fit to $N\propto\frac{1}{\sqrt{\omega_{C,N}}}$ in mode 4. for small values (equation \eqref{eq:N_asinh}) and mode 1. for large values (equation \eqref{eq:N_slow}) is shown as dotted lines. (a) Fast rotation (large rotation frequency $\omega_{C,N}$). (b) Slow rotation
}
\end{figure}

For particles with low $\omega_{C,1}$, which corresponds to almost isotropic particles, where $\frac{\omega_{C,N}}{\omega_{C,1}}>4$, mode 4. are observed. The critical chain length in mode 4. can be well approximated as a function proportional to $\frac{1}{\sqrt{\omega_{C.N}}}$. In Fig.\ref{fig:mode4fit} linear fit is shown for the critical chain length for chains where the critical chain length is in mode 4. In Fig.\ref{fig:mode4fit} for susceptibility values $\chi_\|/\chi_\perp=1.5$ and $\chi_\|=2$ also linear fit in the region $\omega_{C,N}/\omega_{C,1}>1$ where the critical chain is in mode 1. can be observed. The linear fits have different slopes in mode 1. and mode 4. for the same particles. 

Similarly the critical chain length is proportional to $\frac{1}{\sqrt{\omega_{C.N}}}$ in mode 1., where particles have large magnetic anisotropy and $\frac{\omega_{C,N}}{\omega_{C,1}}<1$. Linear fit in mode 1. is shown in Fig. \ref{fig:mode1fit}. These findings agree with the theory from section III.

In modes 2. and 3. the critical chain length deviates from proportionality to $\frac{1}{\sqrt{\omega_{C.N}}}$ as can be seen in Fig. \ref{fig:mode4fit} and Fig. \ref{fig:mode1fit} for $\frac{\omega_{C,N}}{\omega_{C,1}}$ values close to unity. In modes 2. and 3. there can be observed a switch from the slope of mode 1. seen in Fig. \ref{fig:mode1fit} and in Fig. \ref{fig:mode4fit} to the right to the slope of mode 4. seen in Fig. \ref{fig:mode4fit}. Since particles rotate asynchronously between each other, the critical chain length is a bit shorter than in neighbouring modes 1. and 4.

\subsection{Distribution of chain lengths in a small cluster}

If at the beginning the particles are randomly distributed in a limited area and a rotating field is being applied, then particles get organized in planes where distribution of chain lengths can be observed. If $r_{max}$ is smaller than $\sqrt{5}R$, particles create chains with the width larger than 1 and other large structures. The region where $r_{max}<\sqrt{5}R$ is not studied. If $r_{max}$ value is larger than $\sqrt{5}R$, in simulations $r_{max}=2.3 R $, particles create one particle thick chains. Chains rotate next to each other, collide and exchange with particles. As a result, chains of different length can be observed.

The length of each chain is being measured and chain length distribution is obtained. If the distance between two particles is less than $2.1R$, then they are being considered  neighbors. If a particle has one neighbor, then it is an end particle of a chain. If a particle has two neighbors, then it is one of the middle particles of the chain. If a particle has more than two neighbors, then the structure is not a  1 particle thin chain and all the neighboring particles are excluded from the chain distribution results. This gives an algorithm to obtain chain length distribution.

The simulations were done in an infinite box with no boundaries. In most cluster simulations 61 or 91 particles were used. To verify the results, some simulations were done also with 127 and 169 particles (numbers 61,91,127,169 are centred hexagonal numbers). Initially particle centers were randomly distributed in a cylindrical volume of a height $2R$ and a diameter that corresponds to a hexagonal crystal with a lattice constant $8R$, i.e. the diameter of the initial cluster is $80R$ for 61 particles and the diameter of the initial cluster is $96R$ for 91 particles. The tests showed that larger or smaller initial density of particles increased the time to reach a stable histogram.

\begin{figure}[htbp]
\subfigure{\includegraphics
 [width=240pt]
  {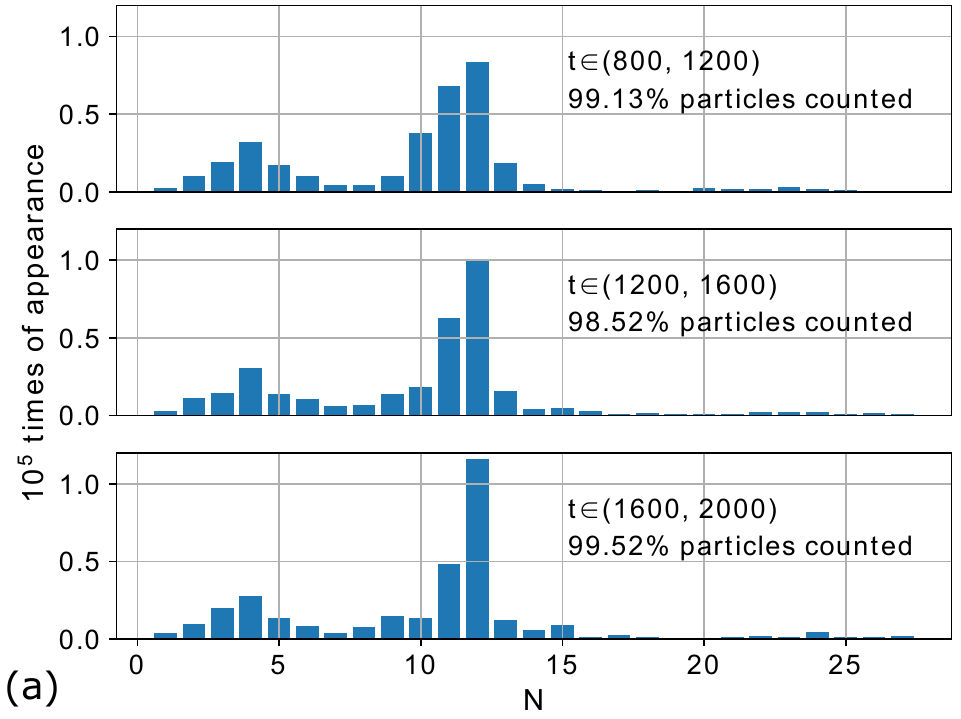} \label{fig:sadalijums_laikaa}}
\subfigure{\includegraphics
 [width=240pt]
  {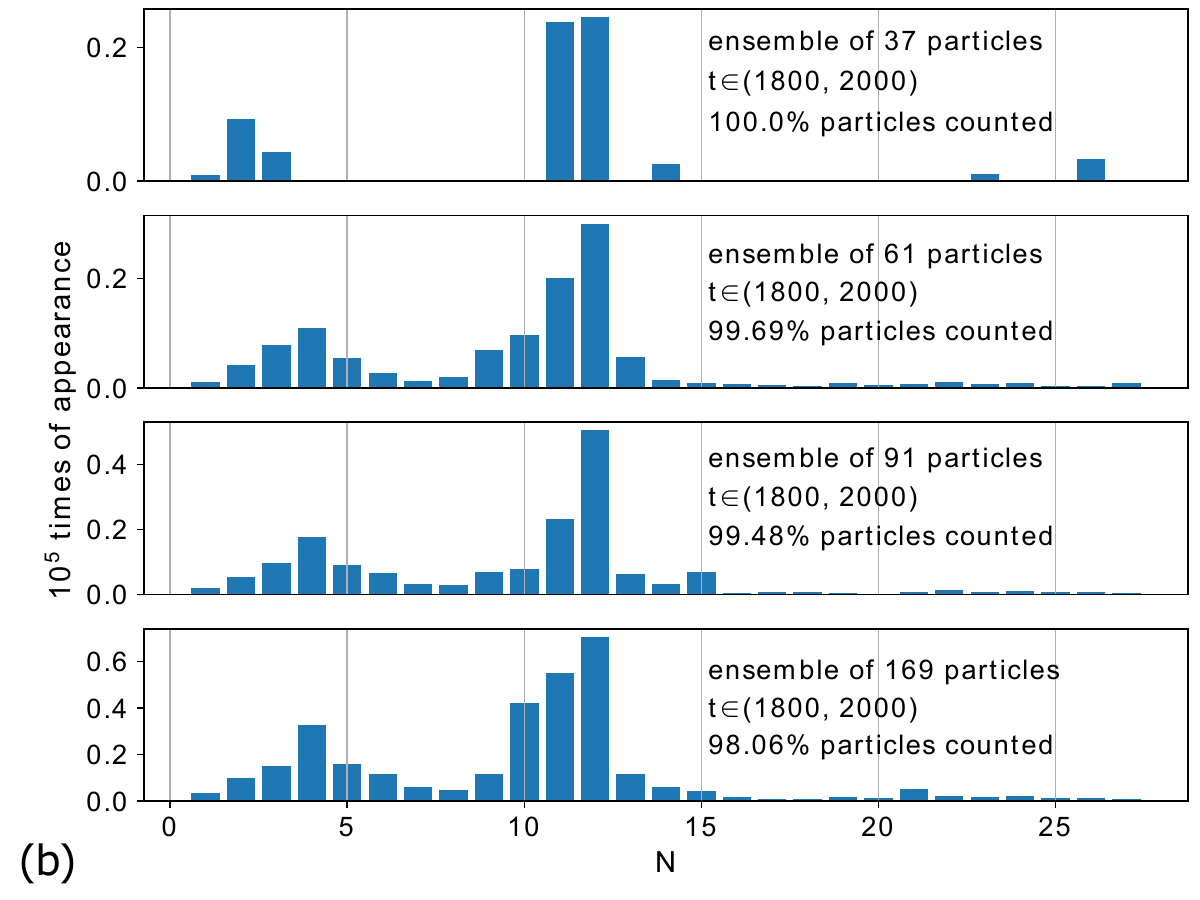} \label{fig:sadalijums_skaits}}
\caption{ (a) The chain length $N$ distribution development over the time in the cluster of 91 particles. (b) The chain length $N$ distribution dependence on the number of particles in the cluster. Simulation gives typical chain length $N_{ensemble}=12 \pm 0.5$. Each histogram contains the sum of chain length distribution histograms time interval $t$ given in the number of periods from the beginning of the calculation and 100 frames per period.
$\frac{\chi_\|}{\chi_\perp}=1.5$ and $\chi_\|=0.25$ at field frequency $\frac{\omega_{H}}{\omega_{C,1}}=0.1$.
}
\end{figure}

The development of the distribution of chain lengths can be seen in Fig. \ref{fig:sadalijums_laikaa} for particle parameters $\frac{\chi_\|}{\chi_\perp}=1.5$ and $\chi_\|=0.25$ at field frequency $\frac{\omega_{H}}{\omega_{C,1}}=0.1$. The histogram of chain lengths is obtained by counting chains over 400 subsequent periods and 100 frames in each period. As can be seen in the histogram, in the first 2000 periods a typical chain length of 12 particles developed.



To simulate an ensemble of particles an infinite number of particles or periodic boundary conditions should be used in simulations. Our simulations show that 2-3 times more particles than the critical chain length is enough to get the distribution of chain lengths in the cluster. If more particles are being used or the initial layout of particles, or particles density are changed then only the time for reaching the stable configuration changes, but the final distribution of chains lengths is not significantly changed. The distribution of chain lengths in a cluster dependence on the initial number of particles is shown in Fig. \ref{fig:sadalijums_skaits}. The histogram of 61 particles is similar to the histograms with higher particle count could be because 61 particles is enough to create all possible chain length configurations in different time steps and averaging over time is similar to averaging over space. 37 particles create three rotating chains which poorly collide with each other.

In mode 1. there is a certain chain length that dominates in the distribution of chains lengths, but in all other modes the width of the dominating peak is larger. 
The error bars are being calculated as a half-width at half of maximum.
 
\begin{figure}[htbp]
\centering
\subfigure{
\includegraphics
 [width=240pt]
  {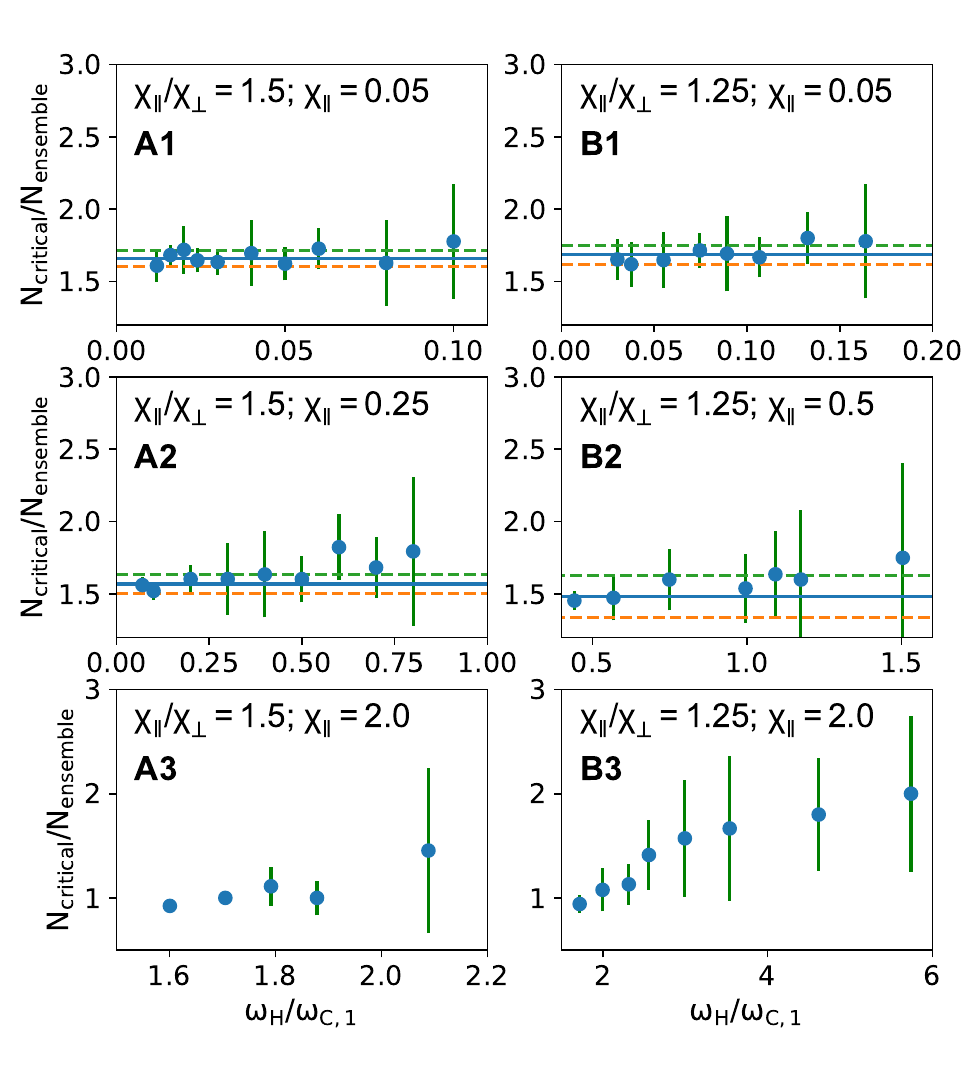}}
\\
\vspace{-0.6cm}
\subfigure{
\includegraphics
 [width=120pt]
  {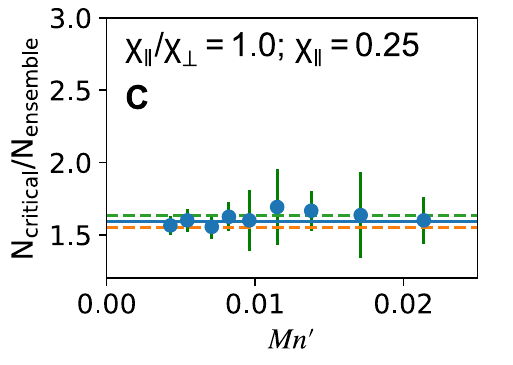}}  
\caption{ \label{fig:NN_grafiki} Ratio of the critical length of one chain and the typical length of particles $\frac{N_{critical}}{N_{ensemble}}$ for particles with different magnetic properties. Weighted mean value of ratio is illustrated as a solid line and weighted standard deviation is illustrated with a dashed line. Error bars represent the width of the distribution of $N$ in a cluster of particles. In graph C, shown ratio $\frac{N_{critical}}{N_{ensemble}}$ for isotropic particles. In graph C, Mason number \eqref{eq:Mn} is used as x-axis. Ratios in graphs A1, A2, B1, B2 and C are constant within standard deviations. Ratios in graphs A1 and B1 coincide within standard deviation. The larger magnetic susceptibility the smaller the ratio is. 
}
\end{figure}

The ratio between the critical chain length of a single chain and the typical chain length in a cluster of chains has been calculated. Numerical simulations show that the critical and the typical chain length ratio is constant for certain type of particle
as can be seen in Fig. \ref{fig:NN_grafiki} graphs A1, A2, B1 and B2.
Ratio mean values coincide within its standard deviations in cases A1 and B1 ($1.65\pm0.06$ for A1 and $1.68\pm0.07$ for B1).
For larger $\chi_\|$ the ratio between the critical and the typical chain length decrease, as can be seen in Fig. \ref{fig:NN_grafiki} graph A2 (mean value $1.57\pm0.07$) and B2 (mean value $1.48 \pm 0.15$).
Graph B2 includes particle motion not only in mode 1., but also modes 2.b and 3.
In this case the ratio does not change much as the modes change.

In figures A3 and B3 Fig. \ref{fig:NN_grafiki} typical chain length in a cluster becomes close to the critical chain length and the ratio becomes one. The ratio becomes one if the critical chain length is in mode 2.a.
The mode 2.a can be observed in A3 and B3 Fig. \ref{fig:NN_grafiki} for  frequencies up to $\frac{\omega_H}{\omega_{C,1}}=1.79$ in graph A3 and up to $\frac{\omega_H}{\omega_{C,1}}=1.72$ in graph B3. If the frequency is increased, the mode 2.a gradually changes to mode 3., and the ratio becomes larger. The explanation for the fact that in mode 2.a typical chain length becomes equal to typical chain length in a cluster could be following: In a cluster, when two identical chains meet with each other, for a moment they form one twice as long chain, in mode 2.a the chain will split in half because only the end particles of these two chains were disorientated. Unlike in all other modes where all the particles have similar disordered state and have more equal probability become the breaking point.

Rotation frequencies in Fig.\ref{fig:NN_grafiki} were chosen to coincide with the critical frequencies which for most of the parameters can be found in Fig.\ref{fig:graph}. There are two reasons why error bars in Fig.\ref{fig:NN_grafiki} are larger for larger frequencies. Firstly, the chain distribution is wider in the asynchronous regime. Secondly, the chains become shorter for large frequencies. The width of the distribution can not be smaller than 1 which gives error 0.5 for $N_{critical}=2$.

In addition, simulations using isotropic particles were done. Since $\omega_{C,1}=0$ for an isotropic particle, field frequency $\omega_H$ was characterized by Mason number:
\begin{equation}\label{eq:Mn}
  Mn'=\dfrac{\omega_H\zeta_r}{\mu_0 V_m H^2}=\frac{\omega_H 8\pi\eta R^3}{\mu_0 V_m H^2}
\end{equation}
We use apostrophe here ($Mn'$) because different Mason number definitions have been seen in literature. Our definition disagrees these definitions: $Mn'=4Mn$ if $Mn$ from \cite{melle2003,Klingenberg2007} is used; $Mn'=Mn/8$ if $Mn$ from \cite{sherman2015} is used.

Summary of the results of a cluster of isotropic particles can be seen in graph C Fig. \ref{fig:NN_grafiki}  for the critical chain lengths from 8 (for larger $Mn'$) till 18 ( for smaller $Mn'$). As can be seen, the ratio $N_{critical}/N_{ensemble}=1.5$ also for isotropic particles. The width of the chain length distribution (error-bars in graph C Fig. \ref{fig:NN_grafiki})  for isotropic particles is similar to that obtained in synchronous regime of anisotropic particles as can be seen in low frequencies in graphs A1, B1 and A2 Fig, \ref{fig:NN_grafiki}.

\section{Conclusions and discussion}
Numerical simulations show that the critical chain length dependence on field frequency for paramagnetic particles with magnetic anisotropy obeys trend observed for isotropic particles if the field frequency is not close to the critical frequency of individual particle $\omega_{C,1}$. For both particles, magnetically isotropic and anisotropic, the critical chain length in rotating field can be approximated as a function inverse proportional to square root of rotation frequency $N\propto\frac{1}{\sqrt{\omega_H}}$. The deviation from this trend for paramagnetic particles with magnetic anisotropy is the highest if rotation frequency is slightly above the critical frequency of individual particle $\omega_{C,1}$. There is a small difference in the proportionality factor in $N\propto\frac{1}{\sqrt{\omega_H}}$ above and below the critical frequency $\omega_{C,1}$. In conclusion, the observation of the critical chain length does not allow us to identify that the particles have magnetic anisotropy if the field frequency is not close to the critical frequency of an individual particle $\omega_{C,1}$.


Simulations show that the typical length in a cluster of chains is smaller than the maximal (the critical) length of the single chain, for the same frequency. Typical chain length is $1.5$ times smaller than the critical chain length. The exception is the situation where only two end particles of the single chain rotate asynchronously with the field, when the field frequency is close to the critical. In this situation, typical chain length is almost equal to the critical chain length. This can be observed if the rotation frequency is slightly higher than the critical frequency of an individual particle $\omega_{C,1}$ and particles have high magnetic susceptibility. As a result, the fact that the chain length in a cluster is smaller than the critical chain length should be considered, when comparing a theory with an experiment.

In low frequency regime, where field rotation frequency $\omega_H$ is smaller than the critical frequency of an individual particle $\omega_{C,1}$, and particles rotate synchronously with the field, distribution of chain lengths in a cluster is narrower than in high frequency range, where $\omega_H>\omega_{C,1}$. All particles, which have anisotropy, will rotate slowly and the chain length distribution will narrow down. Therefore, to create large amounts of paramagnetic chains with equal length it is better to use paramagnetic particles with magnetic anisotropy then magnetically almost isotropic particles. On contrary, fully isotropic particles will have similar width of distribution of chain lengths as highly anisotropic particles.

In this article a small cluster in an infinite medium is used. This allowed the particles to settle in the optimal configuration where attraction of the particles is compensated by the chain breakage. In contrast, if the large ensemble of particles would have been used, the obtained results would depend on the concentration of the particles. For higher particle concentration, as used in magnetorheological fluids, there would not be enough space for the critical chain length to exist and the typical chain length would be smaller than obtained in simulation in a small cluster. In an ensemble also long range interactions play a significant role, which are negligible in a small cluster. The role of particle concentration, long range interactions and temperature in the chain formation and stability in an ensemble of anisotropic particles are left open for further research.

\section{Acknowledgments}
This work was supported by PostDoc Latvia\newline
 [Project no. 1.1.1.2/VIAA/1/16/060].

\bibliography{anisotropy}

\end{document}